\documentclass{article}
\usepackage{blindtext}
\usepackage[a4paper, total={6in, 8in}]{geometry}

\usepackage{graphicx}
\usepackage[subrefformat=parens,labelformat=parens]{subfig}
\graphicspath{{./figures/}}
\usepackage{dcolumn}
\usepackage{bm}
\usepackage[mathlines]{lineno}
\usepackage[export]{adjustbox}
\usepackage{color}
\usepackage{tabularx,array,booktabs}
\usepackage{caption}
\usepackage{mathtools}
\usepackage{amsmath,amssymb}
\usepackage{soul}
\usepackage{lipsum}
\usepackage{varioref}
\usepackage{algorithm,algpseudocode}
\usepackage{ulem}
\usepackage{siunitx}
\usepackage{comment}
\usepackage{upgreek}
\usepackage{url}

\usepackage{calc}
\usepackage{multirow,multicol}
\usepackage{pdfpages}
\usepackage[shortlabels]{enumitem}
\usepackage{rotating}

\newcommand{\bb}[1]{\mathbf{#1}}
\newcommand{\bs}[1]{\boldsymbol{#1}}

\newcommand{\Comsol}{COMSOL Multiphysics\textsuperscript{\textregistered}}

\begin{document}

\title
{Isospectral open cavities and gratings}

\date{}

  \author{Sebastiano Cominelli\thanks{Politecnico di Milano, Department of Mechanical Engineering, Milano, Italy, (sebastiano.cominelli@polimi.it), Corresponding author.}
    \and
    Benjamin Vial\thanks{Department of Mathematics, Imperial College London, London SW7 2AZ, UK (b.vial@imperial.ac.uk).}
    \and
    Sébastien Guenneau \thanks{The Blackett Laboratory, Department of Physics, Imperial College London, London SW7 2AZ, UK (s.guenneau@imperial.ac.uk).}
    \and
    Richard V. Craster\thanks{Department of Mathematics, 
    UMI 2004 Abraham de Moivre-CNRS, 
    Department of Mechanical Engineering, Imperial College London, London SW7 2AZ, UK. 
    (r.craster@imperial.ac.uk). }
  }

\maketitle
\begin{abstract}
\noindent 

Open cavities are often an essential component in the design of ultra-thin subwavelength metasurfaces and a  typical requirement is that cavities have precise, often low frequency, resonances whilst simultaneously being physically compact. To aid this design challenge we develop a methodology to allow isospectral twinning of reference cavities with either smaller or larger ones, enforcing their spectra to coincide so that open resonators are identical in terms of their complex eigenfrequencies. 
For open systems the spectrum is not purely discrete and real, and we pay special attention to the accurate twinning of leaky modes associated with complex valued eigenfrequencies with an imaginary part orders of magnitude lower than the real part. We further consider twinning of 2D gratings, and model these with Floquet-Bloch conditions along one direction and perfectly matched layers in the other one; complex eigenfrequencies of special interest are located in the vicinity of the positive real line and further depend upon the Bloch wavenumber. The isospectral behaviour is illustrated, and quantified, throughout by numerical simulation using finite element analysis.  

\end{abstract}

\maketitle

\section{Introduction}
\label{sec:intro}

Acoustic metasurfaces, which have subwavelength thickness often constructed from local resonating structures, have found wide application to sound absorbing surfaces and surfaces that manipulate sound in often remarkable ways \cite{cummer2016,romerogarcia2020}; the  performance of these surfaces is due to the resonators embedded in the surfaces and the resonance frequencies are directly connected to the geometry and volume of the resonant cavity with low frequencies requiring large volume. Many approaches have been taken to ensure metasurfaces remain ultra thin, even at low frequencies and approaches such as spiral or labyrinthine resonators are popular to save space and give added functionality \cite{liang2012}. An approach, explored here, is to design smaller cavities that have matching wave frequency spectrum to larger reference cavities by designing the material properties within the resonator accordingly. We show that one can go further than simply match a single resonant frequency and one can match (i.e. twin) the entire wave frequency spectrum and hence match the entire behaviour of the cavities at all frequencies. Cavities with this perfect matching are called isospectral cavities and in a mathematical sense are defined as cavities with different shapes that exhibit identical eigenfrequencies, and their design presents an intriguing challenge in the field of wave physics. The ability to design such twinned cavities opens up new possibilities across various disciplines, including acoustics, electromagnetism, or water waves. By achieving isospectrality, two distinct resonators can exhibit indistinguishable wave behaviour, enabling applications such as creating rooms or auditoria with different geometries that possess identical sound characteristics or engineering elastic components with shared vibrational eigenfrequencies.

The history of isospectral problems can be traced back to the question famously posed by Mark Kac regarding whether one can hear the shape of a drum \cite{kacCanOneHear1966}. The investigation of isospectral drums, where the Laplacian operator within closed domains with Dirichlet boundary conditions yields identical spectra for distinct regions sharing the same area, has produced significant results for specific cases and subsets of the problem \cite{gordonOneCannotHear1992}. These isospectral problems in bounded domains are closely linked to inverse problems in open space \cite{sleemanInverseProblemAcoustic1982}, forming a rich area of research in the past. Furthermore, there has been a recent interest in the study of isospectral or quasi-isospectral potentials inspired by supersymmetric transformations as applied to electromagnetism \cite{parkHearingShapeDrum2022}. We proceed to utilise \textit{Transformation Acoustics} (TA), which has been extensively applied for manipulating wave propagation in diverse physical fields sharing the same analytical structure, like electromagnetism \cite{pendry2006controlling}, acoustics \cite{norris2008acoustic}, elasticity \cite{milton2006cloaking,brun2009achieving,norris2011elastic}, and many others fields \cite{kadic2015experiments}. 
Perhaps the most striking and well-studied effect enabled by TA is cloaking, allowing perfect concealment of an arbitrary object using a singular transformation. However, another strategy proposed in the work by Li \textit{et al.\ } \cite{li2008hiding}, the so-called \textit{carpet cloak}, requires non-singular transformations, making this route more suited for the practical implementation of the equivalent properties. Twinning closed cavities through TA has been recently developed by  Lenz~\textit{et al.\ } in \cite{lenz2023transformation}, where the discrete spectrum of a closed domain with Dirichlet boundary conditions is successfully matched. Here  we consider unbounded open domains; this is not straightforward as spectral problems for open cavities  allow for the leakage of energy into the unbounded medium, have complex eigenspectra and further complications for both theoretical and numerical aspects; we are unaware of attempts to achieve isospectral domains in open systems in wave physics and this opens the way to, for instance, twinning optical waveguides. 

\textit{In primis}, the eigenvalue problem of an unbounded open domain gives rise to complex valued eigenfrequencies $\omega$, whose real part gives the resonant frequency and imaginary part describes the radiation losses through the unbounded region. One can distinguish eigenmodes in two categories: on one hand, the ones linked to a localised resonance within the cavity are called \textit{trapped modes} and they are characterised by a high quality factor $Q= \Re{(\omega)}/2|\Im{(\omega)}|$. 
On the other hand, the so-called \textit{leaky modes} or \textit{quasi-normal modes} \cite{hu2009understanding}
exhibit higher damping in time, which gives rise to an exponential divergence in space, making them more challenging to compute numerically \cite{evripides2022}. In practice, it is not clear how sensitive the matching of such modes can be in terms of both real and imaginary part, particularly when those are of different magnitude. 
Moreover, the eigenspectra of open cavities show both continuous and discrete branches \cite{bamberger1990mathematical,zolla2005foundations} and a non-singular transformation is required so that their topology is preserved \cite{greenleaf2009cloaking,kohn2010cloaking}.

Finally, from a numerical standpoint, the computation of resonances in open systems is complicated by the reflection of the leaky waves at the necessarily finite-grid boundaries. To tackle this issue, we use \textit{Perfectly Matched Layers} (PML), whose analysis and performance for spectral problems have been extensively documented in the literature, see, e.g., \cite{hein2004resonances,harari2000analytical, olyslagerDiscretizationContinuousSpectra2004, dhiaFiniteElementComputation2009a, vialQuasimodalExpansionElectromagnetic2014}.

The manuscript is organised as follows: Section~\ref{sec:motivation} briefly outlines the main transformation acoustics tools and the notation adopted. In order to highlight the main effects and underlying ideas of the twinning process in a simple setting, we then give the closed form solution for two one-dimensional domains. 
In Section~\ref{sec:num sol}, we study the spectrum of an open cavity whose geometry allows both leaky and resonant modes. The transformed region is then further discretised using a layered medium that approximates the required anisotropic properties to demonstrate that conceptually one can use simpler building blocks to achieve the desired effects. 
Section~\ref{sec:metasurf} extends the method to the periodic case: we show that a saw-tooth grating can be matched by a flat, thinner metasurface by comparing the dispersion diagrams for the original and transformed elementary cell. 
Finally, in Section~\ref{sec:conclusions} we draw some conclusions and discuss potential extensions of the present work.

\section{Twinning through transformation}
\label{sec:motivation}

We take a domain filled with an acoustic fluid supporting a pressure field, for time-harmonic waves, that satisfies Helmholtz's equation, where the filling fluid is characterised by a density $\rho_0$ and a bulk modulus $\kappa_0$. The same procedure can be extended to other areas of wave physics sharing the same analytical structure i.e.\ {anti-plane} shear waves in elasticity and polarised waves in electromagnetism; we choose to use the setting of pressure acoustics without loss of generality. \\
Transformation Acoustics \cite{pendry2006controlling,milton2006cloaking,norris2008acoustic} shows how to project a subdomain $\Sigma_c$  into a deformed subdomain $\sigma_c$ using the map $\chi\colon\bb X\mapsto\bb x$ while  preserving the acoustic behaviour in the remaining subdomains, if particular properties are chosen inside $\sigma_c$, and we follow this approach here.
We indicate with capital letters $\bb X$, $P$, and $\Omega$ the set of coordinates, the pressure field, and the frequencies respectively, defined on the undeformed (or material) domain $\Sigma$; lowercase letters $\bb x$, $p$, and $\omega$ that refer to the deformed (or spatial) domain $\sigma$.


The generalised eigenvalue problem
\begin{equation}
    -\nabla_{\bb X}\cdot(\rho_0^{-1}\nabla_{\bb X} P) = \Omega^2\kappa_0^{-1} P,
\end{equation}
written with respect to the
material domain $\Sigma$, is equivalent to
\begin{equation}\label{eq:transf helmholtz}
    -\nabla_{\bb x}\cdot(\bs\rho^{-1}_c\nabla_{\bb x} p) = \omega^2\kappa^{-1}_c p,
\end{equation}
that describes the pressure $p$ with respect to the
spatial domain $\sigma$. Here, $\bs\rho_c$ is a tensor describing the equivalent anisotropic density and $\kappa_c$ is the equivalent bulk modulus after transformation. They are defined as
\begin{align}\label{eq:transf prop}
    \bs\rho_c &\coloneqq \rho_0\bb J \bb J^\top /\det{\bb J},
    &
    \kappa_c&\coloneqq\kappa_0\det{\bb J},
\end{align}
where $\bb J\coloneqq \partial_{\bb X}\bb \chi(\bb X)$ is the Jacobian matrix associated with the transformation (a push-forward) that describes the local deformation of the geometry. In the following we use a non-singular transformation to define two open cavities with different shapes but sharing the same spectrum.

\subsection{1D analytical example: twinning a semi-infinite pipe}
\label{sec:motivation-a}
\begin{figure}
    \centering
    \subfloat[]{\includegraphics[width=0.45\textwidth,trim=20 -33 20 0]{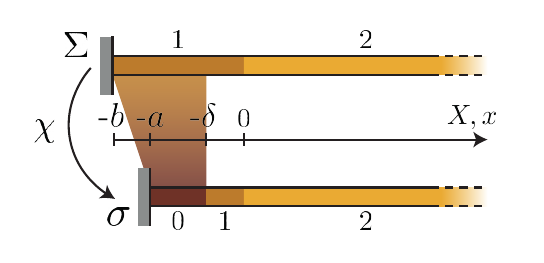}\label{fig:1d shortened-a}}
    \quad
    \subfloat[]{\includegraphics[width=0.5\textwidth,trim = 0 30 0 0]{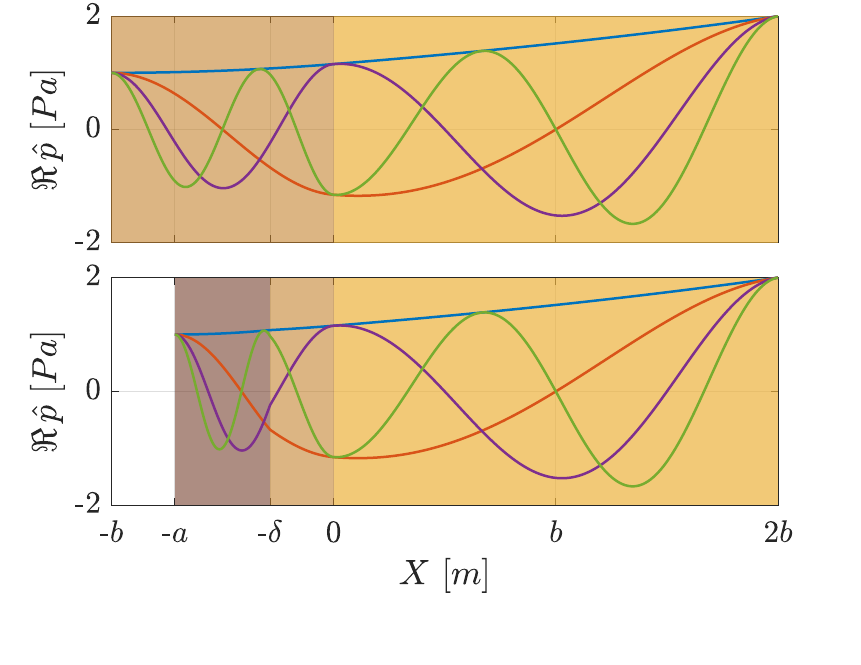}\label{fig:1d shortened-b}}
    \caption{\protect\subref{fig:1d shortened-a} A schematic of the reference leaky resonant cavity (top) with domain $\Sigma$ and the transformed (bottom) domain $\sigma$. \protect\subref{fig:1d shortened-b} A plot of the real part of the first four eigenmodes: (top) the reference configuration, (bottom) the transformed one. The modes inside $[-b,-\delta]$ of the reference configuration are compressed into $[-a,-\delta]$ for the transformed domain whilst they match for $[-\delta,+\infty)$.
    }
    \label{fig:1d shortened}
\end{figure}

We outline the implications of this approach by considering a simple one dimensional problem that admits a closed form solution.\\
Let us consider the 1D semi-infinite spaces $\Sigma$ and $\sigma$ as shown in Figure~\ref{fig:1d shortened-a}, they contain either one or two slabs of fluids different from the matrix. Namely, fluid $i$ is characterised by a density $\rho_i\in\mathbb{R}$ and bulk modulus $\kappa_i\in\mathbb{R}$, and it is contained in the geometric interval $\Sigma_i$ (or $\sigma_i$), for $i=\{0,1,2\}$. Note that a domain with only one slab can be easily handled as the special case of two slabs having the same properties. As a consequence, we can solve the eigenvalue problem for the most general configuration with two slabs and then readily analyse the pipe with only one slab. Note that this configuration is of interest because the slabs, having acoustic properties different from the matrix, behave like a cavity: this similarity leads to eigenmodes that are partially trapped inside the slab and that leak energy through the unbounded matrix.

Let us consider the domain $\sigma$ in Figure~\ref{fig:1d shortened-a}. The eigenvalue problem describing the complex valued pressure $p(x)\colon\mathbb R\to\mathbb C$ on the three acoustic domains reads as:
\begin{equation}
\begin{aligned}
        -\partial_{xx} p_i(x) &= \frac{\omega^2}{c_i^2} p_i(x),& x\in \sigma_i,\, i=\{1,2,3\}
\end{aligned}
\end{equation}
where we introduce the sound speed $c_i=\sqrt{\kappa_i/\rho_i}$, and the intervals $\sigma_0\coloneqq[-a,-\delta)$, $\sigma_1\coloneqq[-\delta,0)$ and $\sigma_2\coloneqq[0,+\infty)$, for $i=\{1,2,3\}$.
The boundary value problem is completed by the rigid boundary condition on the left end and the Sommerfeld condition at infinity
\begin{equation}
\begin{aligned}
        &-\partial_x p_0(-a)=0,
        &
        \lim_{x\to+\infty} &\left(\partial_x p_2(x) + jk_2p_2(x) \right)= 0,
\end{aligned}
\end{equation}
where we have adopted the time-harmonic convention $e^{j\omega t}$, being $j$ the imaginary unit, i.e.\ ${j^2=-1}$.
Moreover, the equilibrium and compatibility conditions hold on the interfaces at $x=-\delta$ and $x=0$ 
\begin{equation}
\begin{aligned}
        p_0(-\delta^-)&=p_1(-\delta^+), 
        &
        \frac{1}{\rho_0}\partial_x p_0(-\delta^-) &= \frac{1}{\rho_1}\partial_x p_2(-\delta^+),
        \\
        p_1(0^-)&=p_2(0^+),
        &
        \frac{1}{\rho_1}\partial_x p_1(0^-) &= \frac{1}{\rho_2}\partial_x p_2(0^+).
\end{aligned}
\end{equation}
By considering the dispersion relations $\omega=k_ic_i$, $i=\{0,1,2\}$ and choosing the \textit{ansatz}s $p_0(x)=A_0\cos(k_0x+\varphi_0)$, $p_1(x)=A_1\cos(k_1x+\varphi_1)$ that account for the resonances trapped in the slabs and $p_2(x)=A_2e^{-jk_2x} + B_2e^{+jk_2x}$ for the leakage through the matrix, the eigenvalue problem is solved if the boundary conditions are applied. We thus obtain the following equalities:
\begin{equation}\label{eq:1d equalities}
    \begin{aligned}
        \varphi_0&=k_0a,
        &
        A_0\cos{\big(k_0(a-\delta)\big)}&=A_1\cos{(\varphi_1-k_1\delta)},
        &
        A_1\cos(k_1b)&=A_2,
        \\
        B_2&=0,
        &
        \frac{k_0}{\rho_0}\tan{\big(k_0(a-\delta)\big)} &= \frac{k_1}{\rho_1}\tan{(\varphi_1-k_1\delta)},
        &
        \frac{k_1}{\rho_1}\tan{\varphi_1} &= -j\frac{k_2}{\rho_2}.
    \end{aligned}
\end{equation}

If $\rho_0=\rho_1$ and $\kappa_0=\kappa_1$, the two slabs merge together and the domain $\Sigma$ shown in Figure~\ref{fig:1d shortened-a} can be considered. In that case, equalities~\eqref{eq:1d equalities} simplify to
\begin{equation}
\begin{aligned}
    \varphi_0=\varphi_1&=k_1b,
    &
    A_0&=A_1,
    &
    B_2&=0,
    &
    A_2&=A_1\cos(k_1b),
    &
    \frac{k_1}{\rho_1}\tan{(k_1b)} &= -j\frac{k_2}{\rho_2}.
\end{aligned}
\end{equation}
This defines the set of eigenvalues
\begin{align}\label{eq:1D eigf}
    \Omega^{(n)} &=  \frac{ c_1}{b}\left[  n \pi + \frac{j}{2}\ln{\left(\frac{z_2+z_1}{z_2-z_1}\right)} \right] , & n\in\mathbb Z
\end{align}
where $z_i=\sqrt{\kappa_i\rho_i}$ is the impedance of medium $i$ 
and $\Omega^{(n)}=K_1^{(n)}c_1 = K_2^{(n)}c_2$. In the field of optics, this corresponds to Fabry-Perot resonances \cite{Perot1899, Ismail2016} at which light exhibits constructive interference after one round trip, 
with the imaginary part giving the the lifetime of a photon inside the cavity $\tau_n = 2\pi/\Im\Omega^{(n)}$. 
The corresponding eigenmodes are given by
\begin{equation}
\begin{aligned}
    \hat P_1^{(n)} &= \cos{\big(K_1^{(n)}(X+b)\big)},
    &
    \hat P_2^{(n)} &= \cos{(K_1^{(n)}b)}\,e^{-jK_2^{(n)}X}.
\end{aligned}
\end{equation}
Note that, since the one slab configuration is referred as the material frame, capital letters are adopted. 
Figure~\ref{fig:1d shortened-b} shows the shape of the first four eigenmodes, where one can remark the exponential growth which is an evidence of the leakage.

In the following, two twins of the pipe are obtained first by compressing a portion of the slab~$1$ and then by folding the space at the extremity of the pipe; both approaches give rise to a second slab of fluid, referred to with the index $0$.
We then compute the properties of the fluid filling the transformed domain $\sigma$ such that the cavities share the same eigenvalues and eigenvectors. 

Let us choose
a linear 
map $\chi: \Sigma\to\sigma, X\mapsto x$ such that $\Sigma_c\coloneqq[-b,-\delta]\to\sigma_c\coloneqq[-a,-\delta]$ and $\Sigma_f\coloneqq(-\delta,+\infty)\to\sigma_f\coloneqq(-\delta,+\infty)$. 
That is
\begin{equation}
    \chi(X)\coloneqq
    \begin{dcases}
        \frac{a-\delta}{b-\delta}X + \delta\frac{a-b}{b-\delta}, & \text{if } X\in [-b;-\delta]
        \\
        X,& \text{if } X>-\delta
    \end{dcases}
\end{equation}
In this case, the deformation gradient $J=\partial_X x
=\frac{a-\delta}{b-\delta}
$ is a scalar, so we obtain the following transformed properties
\begin{align}\label{eq:prop changes}
    \rho_0&=\frac{b-\delta}{a-\delta}\rho_1, &
    \kappa_0&=\frac{a-\delta}{b-\delta}\kappa_1, &
    c_0&=\frac{a-\delta}{b-\delta}c_1,  &
    z_0&= z_1
\end{align}
by usage of the formulae~\eqref{eq:transf prop}. The resulting acoustic system is the one of a shorter pipe having two slabs of homogeneous properties, depicted in Figure~\ref{fig:1d shortened-a}.
Its eigenmodes are readily computed through equalities~\eqref{eq:1d equalities}:
\begin{align}
    \hat p_0^{(n)} &
    = \cos{\big(k_0^{(n)}(x +a)\big)},
    &
    \hat p_1^{(n)} &= \cos{\big(k_1^{(n)}(x+b)\big)},
    &
    \hat p_2^{(n)} &= \cos{(k_1^{(n)}b)}\,e^{-jk_2^{(n)}x},
\end{align}
where $k_0^{(n)}=J^{-1}K_1^{(n)}$, $k_1^{(n)}=K_1^{(n)}$, and the eigenvalues $\omega^{(n)}=\Omega^{(n)}$ are the same of the original cavity given by Eq.~\eqref{eq:1D eigf}.
Note that the eigenmodes are such that $\hat p^{(n)}(x) = \hat P^{(n)}\big(\chi^{-1}(x)\big)$.

As just illustrated, it is possible to transform the space and obtain a shorter or a longer twin cavity. However, as highlighted by Eq.~\eqref{eq:prop changes}, the properties change proportionally to the geometric stretch: if $\rho$ is increased, $\kappa$ should decrease accordingly. In general, a material with such properties is not readily attainable, so in the following we define a slightly different strategy for twinning a cavity relying on space folding.

\begin{figure}
    \centering
    \subfloat[]{\includegraphics[width=0.45\textwidth,trim=20 -33 20 0]{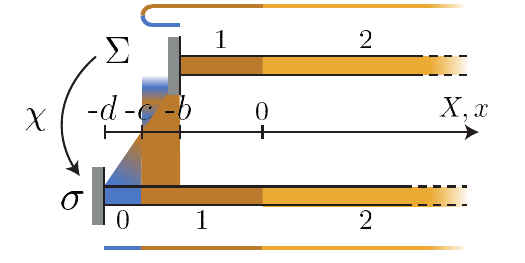}\label{fig:1d elongated-a}}
    \quad
    \subfloat[]{\includegraphics[width=0.5\textwidth,trim= 0 30 0 0]{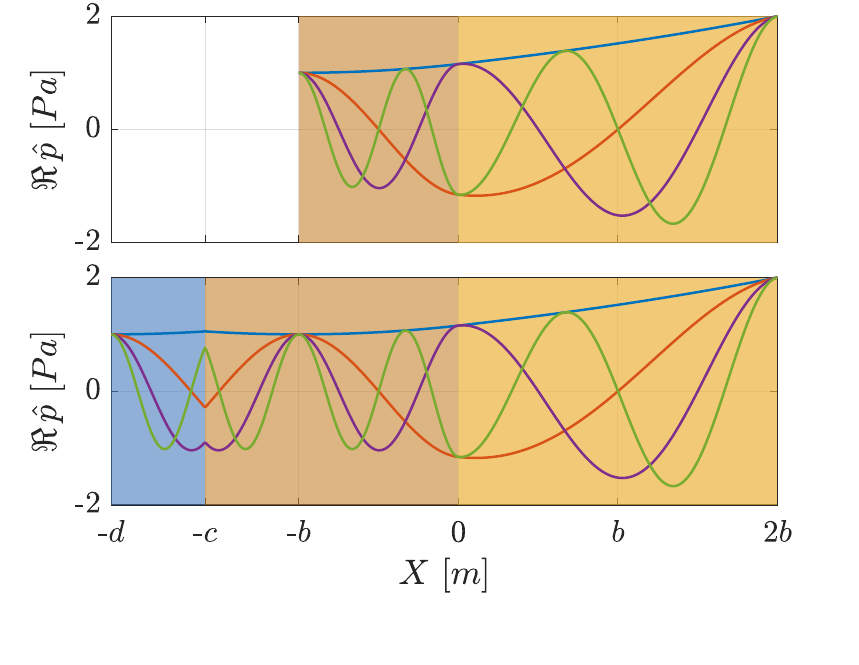}\label{fig:1d elongated-b}}
    \caption{\protect\subref{fig:1d elongated-a} A schematic of the reference leaky resonant cavity (top) and the transformed (bottom) domain. \protect\subref{fig:1d elongated-b} The plot of the real part of the first four eigenmodes: (top) the reference configuration, and (bottom) the transformed one. The modes inside $[-d,-b]$ of the transformed configuration are symmetric with respect to the point $-c$ and, if folded into $[-c,-b]$, they annihilate each other; the eigenmodes are preserved in the half space $\left[-b,+\infty\right)$.
    }
    \label{fig:1d elongated}
\end{figure}

It is well known that space folding transformations \cite{leonhardt2006general} lead to Negative Index Materials (NIM) \cite{veselago1967electrodynamics,shelby2001experimental,pendry2003focusing,ramakrishna2005physics,lai2009complementary,ramakrishna2004spherical}, whose effective properties are the consequence of localised resonances. Hence, we can leverage this mathematical abstraction to unlock alternative transformations and create another route for the design of isospectral open cavities.\\
Let us consider the schematic on top of Figure~\ref{fig:1d elongated-a}, the space $[-c,-b)$ beyond the rigid extremity of the pipe is considered as the overlap of two complementary materials \cite{pendry2003focusing} whose acoustic behaviours annihilate each other.
Through the transformation $\chi$, such that
\begin{equation}
    X = \chi^{-1}(x) =
    \begin{dcases}
        x, & x\ge -c
        \\
        -x - 2c, & -d\le x<-c
    \end{dcases}
\end{equation}
they are unfolded onto a straight line and, like the negative of a photograph, their properties are equal but with opposite signs: $\rho_0=-\rho_1$, $\kappa_0=-\kappa_1$ (doubly negative acoustic parameters can be achieved in practice with single resonance metamaterials \cite{li2004double}, that constrains twinning to a small portion of the spectrum).
This arrangement is highlighted in Figure~\ref{fig:1d elongated} using complementary colours. Note that we chose $c=(b+d)/2$ such that only unitary (positive or negative) stretches arise, however different choices are possible.\\
By applying once again the equalities~\eqref{eq:1d equalities}, the closed form of the twin cavity eigenvalues is achieved:
\begin{align}
    \hat p_0^{(n)} &= \cos{\big(k_0^{(n)}(x+d)\big)},
    &
    \hat p_1^{(n)} &= \cos{\big(k_1^{(n)}(x+b)\big)},
    &
    \hat p_2^{(n)} &= \cos{(k_1^{(n)}b)}\,e^{-jk_2^{(n)}x},
\end{align}
where the wavenumbers $k_i^{(n)}$ are obtained using the dispersion relation $\omega_i^{(n)}=k_i^{(n)}c_i$, and the eigenvalues are the same as Eq.~\eqref{eq:1D eigf}; thus the cavity is a twin.

These trivial 1D cases share many features with several practical applications, for instance a duct where pressure waves propagate, or a string under tension sustaining transverse waves, etc. One can therefore substitute a portion of the domain with a shorter/longer one with equivalent parameters whilst keeping the same eigenfrequencies and alter the eigenmodes inside the transformed region only.

\section{Manipulating a Helmholtz resonator}
\label{sec:num sol}
\begin{figure}
    \centering
    \includegraphics[width=0.85\textwidth,trim=0 10 0 5]{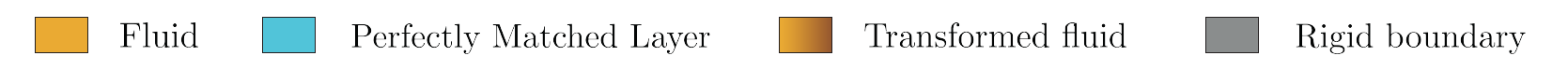}
    \\
    \subfloat[]{\includegraphics[height=4.5cm]{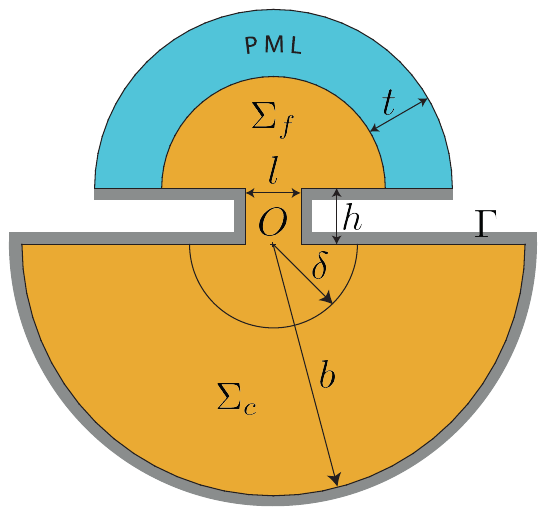} \label{fig:geometry-a}}
    \subfloat[]{\includegraphics[height=4.5cm,trim=10 0 25 0]{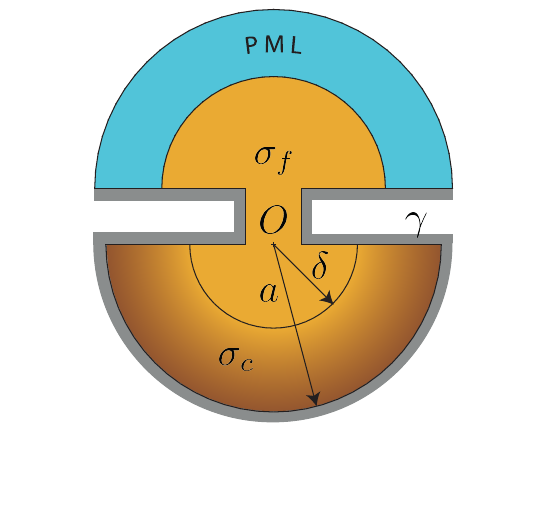} \label{fig:geometry-b}}
    \subfloat[]{\includegraphics[height=4.5cm,trim=20 0 0 0]{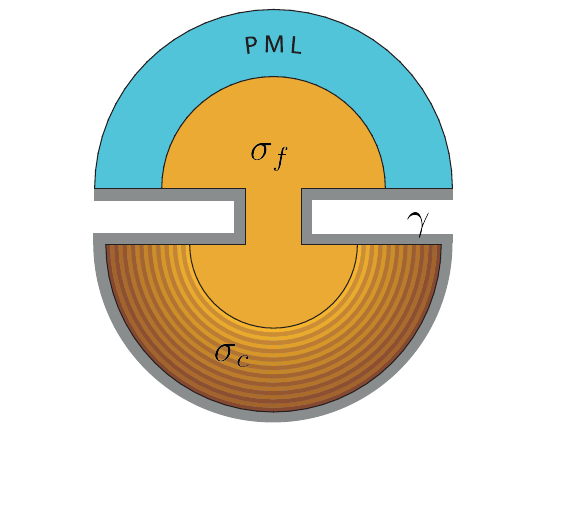} \label{fig:geometry-c}}
    \caption{\protect\subref{fig:geometry-a} Virtual geometry of the reference cavity, both $\Sigma_f$ and $\Sigma_c$ are filled with a homogeneous fluid; \protect\subref{fig:geometry-b} ideally twinned cavity, $\sigma_f$ is filled with a homogeneous fluid while $\sigma_c$ is filled with the anisotropic fluid; \protect\subref{fig:geometry-c} the anisotropic fluid is approximated by a graded layered metamaterial. The geometry is defined by $h=l=\SI{1}{\meter}$, $t=1.2l$, $a=3l$, $b=1.5a$ and $\delta=0.5a$.
    }
    \label{fig:geometry}
\end{figure}

We now consider a 2D open cavity and cross-validate computations of its spectrum against numerical simulations of an ideal anisotropic fluid and also against an effective medium using layers \cite{torrent2008anisotropic}, each having constant material properties, which illustrates an approach towards experimental realization. 
Let us consider the geometry depicted in Figure~\ref{fig:geometry-a}, the cavity is assumed to have sound hard boundary $\Gamma$ and it communicates with the  exterior medium through a small aperture; an artificial boundary is introduced to truncate the computational domain and it is surrounded by a \textit{Perfectly Matched Layer} (PML) that mimics an infinite domain, see, e.g., \cite{hein2004resonances}.
The cavity is large compared to the aperture such that trapped modes are encouraged to arise, while the small aperture allows energy to leak out from the system.


The cavity is filled by a fluid of density $\rho_0=\qty{1}{\kilo\gram\per\cubic\meter}$ and bulk modulus $\kappa_0=\qty{1}{\pascal}$; the computational domain $\Sigma\subset\mathbb R^2$ is partitioned in two non-overlapping sets $\Sigma_f$ and $\Sigma_c$ such that $\Sigma=\Sigma_f\cup\Sigma_c$.
The Helmholtz's equation is supplied with Neumann boundary conditions:
\begin{equation}
    \begin{dcases}
        -\nabla_{\bb X}\cdot\big(\rho_0^{-1} \nabla_{\bb X} P(\bb X)\big)=\Omega^2 \kappa_0^{-1} P(\bb X)& \text{in } \Sigma
        \\
        -\rho_0^{-1}\nabla_{\bb X} P(\bb X)\cdot\bb N=0&\text{on } \Gamma
    \end{dcases}
\end{equation}
where $P(\bf X)$ is the complex valued pressure defined on the material domain $\Sigma$. Note that the eigenvalue problem is obtained considering time-harmonic solutions of the form $\hat P(\bb X,t)=\Re(P(\bb X)e^{j\omega t})$.

A smaller twin cavity is defined by applying Transformation Acoustics, see Section \ref{sec:motivation}, and defining a portion of anisotropic fluid, similar to a  \textit{carpet cloak} \cite{li2008hiding}. Note that a smaller cavity simply filled with the ambient fluid would have higher resonance frequencies. \\
Figure~\ref{fig:geometry-a}-\subref*{fig:geometry-b} illustrates the geometry we consider and the result of the transformation $\bs\chi\colon\Sigma\to\sigma$ that maps the cavity geometry into a smaller one. In particular, the circular annulus $\Sigma_c$ of radii $\delta$ and $b$ is mapped to a smaller one of radii $\delta$ and $a<b$, referred as the deformed domain $\sigma_c$. The transformation is the identity inside the remaining domain $\Sigma_f$, that is simply mapped into $\sigma_f\equiv\Sigma_f$. Also note that the boundary $\Gamma$ is smoothly mapped onto $\gamma$. Figure~\ref{fig:geometry-c} shows the transformed medium with the anisotropic fluid replaced by an effective medium made from layers of homogeneous material. 

Since the cavity is semi-circular, it is convenient to transform the geometry adopting a polar reference system centred in $O$ and described by the set of coordinates $(R,\Theta)$. Then $\bs\chi\colon (R,\Theta)\mapsto(r,\theta)$ is such that
\begin{align}
    r &= f^{-1}(R), & \theta&=\Theta,
\end{align}
where the only requirement is that $f(a)=b$ and $f(\delta)=\delta$. 
We emphasise here that $f(r)$ does not have to be monotonic, meaning situations implying space folding are encapsulated by the present method. This would require \textit{Negative Index Materials} (NIM) \cite{veselago1967electrodynamics,shelby2001experimental,lai2009complementary} whose properties can be approximated by locally resonant material. For the sake of clarity, the effect of such a choice is discussed at the end of this section.
The transformation gradient $\bb J\coloneqq \nabla_{\bb X} \bb x$ is
\begin{align}
    \bb J &=
    \begin{bmatrix}
        \frac{1}{f'(r)} & 0
        \\
        0 & \frac{r}{f(r)}
    \end{bmatrix},
    &
    \det{\bb J}& = \frac{r}{f'(r)f(r)}.
\end{align}
Note that the mixed tensor $\bb J$ is expressed with respect to the canonical contravariant base $(\bb e_r,\bb e_\theta)$ and covariant base $(\bb E^R,\bb E^\Theta)$. Note also that no rotation is implied by the transformation, hence the deformation gradient is symmetric. This turns out to be useful if, for instance, the anisotropic transformed fluid is attained through a pentamode material, the reader is referred to~\cite{norris2008acoustic} for a complete discussion.

\subsection{Twinning via monotonic transform}

For simplicity, we choose the linear function 
\begin{equation}
    f(r) = \frac{b-\delta}{a-\delta}r - \delta\frac{b-a}{a-\delta}
\end{equation}
so the bulk modulus $\kappa_c$ and the tensor density $\bs\rho_c$ of the transformed domain $\sigma_c$ are computed according to formulae~\eqref{eq:transf prop}
\begin{align}
    \bs\rho_c &=
    \rho_0
    \begin{bmatrix}
        \frac{rf'(r)}{f(r)} & 0
        \\
        0 & \frac{f(r)}{rf'(r)}
    \end{bmatrix}
    =
    \begin{bmatrix}
        \rho_r(r) & 0
        \\
        0 & \rho_\theta(r)
    \end{bmatrix},
    &
    \kappa_c&=
    \kappa_0\frac{r}{f'(r)f(r)}
\end{align}
and, because of the axisymmetric geometry, these properties depend on the radial coordinate only.

In view of a real implementation of the anisotropic fluid through a metamaterial, it is fundamental to discuss how it affects the spectrum of the cavity.
The stretch of the geometry induces an anisotropic density whose principal components are aligned with the radial and the tangential directions of the cavity, similarly to the invisibility cloak for axisymmetric obstacles see, e.g., \cite{antonakakis2014gratings}. This situation lends itself to the use of a layered arrangement of two homogeneous and isotropic fluids for achieving the effective properties we need. The following analytical relations hold \cite{antonakakis2014gratings} for the effective properties:
\begin{align}
    \bs\rho_{\rm eff} &=
    \begin{bmatrix}
        \langle\rho\rangle & 0
        \\
        0 & \langle\frac{1}{\rho}\rangle^{-1}
    \end{bmatrix},
    &
    \kappa_{\rm eff}&= \langle1/\kappa\rangle^{-1}.
\end{align}
For our illustration, the anisotropic fluid is discretised by 10 layers of the same thickness $\delta_r=(a-\delta)/10$ each one composed by two materials A and B of thickness $\varepsilon=\delta_r/2$, such that they approximate the anisotropic medium if homogenised two by two. A schematic of such arrangement is shown in Figure~\ref{fig:geometry-c}.

We set $ r^i\coloneqq(2i+1)\varepsilon$, $i\in\{0,\dots,9\}$ the discrete radii pointing in the middle of each layer and $\kappa^i\coloneqq\kappa_c(r^i)$, $\rho_r^i\coloneqq\rho_r(r^i)$, and $\rho_\theta^i\coloneqq\rho_\theta(r^i)$ the transformed properties evaluated at $r^i$. The two materials $A$ and $B$ in the $i$\textsuperscript{th} layer are chosen to have the same bulk modulus such that
\begin{align}
    \kappa_A^i=\kappa_B^i=\kappa^i,
\end{align}
while the densities are chosen as
\begin{align}
    \rho_A^i &= \rho^i_r
    -\sqrt{(\rho_r^i)^2 - \rho^i_\theta \rho^i_r},
    &
    \rho_B^i &= \rho^i_r + \sqrt{(\rho_r^i)^2 - \rho^i_\theta \rho^i_r};
\end{align}
one can easily verify that
\begin{align}
    \langle\rho\rangle=
    \frac{\rho_A^i + \rho_B^i}{2} &= \rho^i_{r},
    &
    \Big\langle\frac{1}{\rho}\Big\rangle^{-1}=
    2\left(\frac{1}{\rho_A^i} + \frac{1}{\rho_B^i}\right)^{-1}
    &= \rho^i_\theta.
\end{align}
Table~\ref{tab:layers} describes the properties of each layer used in the discretized version of the twinned cavity.

\begin{table}[]
    \centering
    $\begin{array}{c|cccccccccc}
    i & 0&1&2&3&4&5&6&7&8&9
    \\    \toprule
    \kappa_A^i=\kappa_B^i \,[\si{\pascal}]& 0.331 & 0.326 & 0.321 & 0.318 & 0.314 & 0.311 & 0.308 & 0.306 & 0.303 & 0.301
    \\ 
    \rho_A^i \quad[\si{\kilo\gram\per\cubic\meter}] & 2.188 & 2.138 & 2.094 & 2.054 & 2.017 & 1.984 & 1.953 & 1.925 & 1.899 & 1.875
    \\ 
    \rho_B^i \quad[\si{\kilo\gram\per\cubic\meter}] & 0.457 & 0.468 & 0.478 & 0.487 & 0.496 & 0.504 & 0.512 & 0.520 & 0.527 & 0.533
    \end{array}$
    \caption{Bulk moduli and densities of each layer composing the metamaterial inside the cavity.}
    \label{tab:layers}
\end{table}

\subsection{Twinning via non-monotonic transform}

Lastly, we briefly consider the transformation enforced by a non-monotonous function $f(r)$ that, as anticipated Section~\ref{sec:motivation}, leads to space folding. To show that the spectrum is matched, let us consider the annular region beyond the curved boundary of the cavity as an overlapped portion of space that can be unfolded into a plane space filled with complementary materials. As suggested by~\cite{lai2009complementary}, the function 
\begin{equation}
    f(r)=
    \begin{cases}
        r, & r\le c \\
        c^2/r, &  r > c
    \end{cases}
\end{equation}
is a natural choice for this circular geometry since it leads to an isotropic yet inhomogeneous medium, whose properties are $\rho_\theta=\rho_r=-\rho_0$ and $\kappa_c=-r^4/c^4$. Where $c=\sqrt{bd}$ and $d=2a$ describe the depth of the twin cavity equipped with a NIM layer, as for the 1D arrangement of section \ref{sec:motivation-a}.

\subsection{Twinning efficiency via modal assurance criterion}

We now turn to a quantitative comparison of the spectrum of the reference cavity with the three configurations just defined (ideal, layer and NIM) and these are all 
 modelled using a finite element method (\Comsol{}). To ensure a fair comparison between the cases, the mesh characteristic size is set to $h<\delta_r/4$ in order to account for the spatially varying properties in the transformed domain and pressure fluctuations, for higher order modes in particular. In addition, the finite element problems share the same mesh in order to minimise the numerical errors due to the discretization.\\
The accuracy of the matching between the eigenfrequencies is expected to deteriorate for increasing frequencies, thus the $20$ eigenvalues with real part larger than and closer to $c_0/5l=\SI{0.2}{\hertz}$ are computed in order to show this trend. Table~\ref{tab:freq comparison} shows their values and highlights the good agreement between the reference and the ideal cases, and a strong match with the layered and NIM configurations. Please note that, in order to improve the numerical convergence for the NIM cavity, a small fictitious damping has been added to the negative material properties such that $\rho_c\rightarrow(1+i\,\num{1e-4})\rho_c$ and $\kappa_c\rightarrow(1+i\,\num{1e-4})\kappa_c$; 
this inherently leads to small differences in the solution of the eigenvalue problem for that case.

\begin{table}
\sisetup{output-exponent-marker=\ensuremath{\mathrm{e}}}
 \centering
$\begin{array}{c|cccc}
\toprule
\text{Mode} &f_{ref}\,\,[\unit{\hertz}] &   f_{ideal}\,\,[\unit{\hertz}]  &  f_{layer}\,\,[\unit{\hertz}] &  f_{NIM}\,\,[\unit{\hertz}] \\ 
\midrule
1&{0.2269 +8.0 i\,\num{e -15}}&{0.2269 +4.9 i\,\num{e -15}}&{0.2219 +5.8 i\,\num{e -15}}&{0.2269 +1.3 i\,\num{e -6}}\\ 
2&{0.2371 +0.0170 i}&{0.2371 +0.0170 i}&{0.2389 +0.0179 i}&{0.2371 +0.0170 i}\\ 
3&{0.2373 +1.5 i\,\num{e -5}}&{0.2373 +1.5 i\,\num{e -5}}&{0.2377 +1.8 i\,\num{e -5}}&{0.2373 +1.6 i\,\num{e -5}}\\ 
4&{0.2653 +5.6 i\,\num{e -14}}&{0.2653 +5.4 i\,\num{e -14}}&{0.2587 +6.5 i\,\num{e -15}}&{0.2653 +1.1 i\,\num{e -6}}\\ 
5&\textcolor{red}{0.2798 +0.1049 i}&\textcolor{red}{0.2798 +0.1049 i}&\textcolor{red}{0.2798 +0.1049 i}&\textcolor{red}{0.2798 +0.1049 i}\\ 
6&{0.2835 +8.1 i\,\num{e -10}}&{0.2835 +8.1 i\,\num{e -10}}&{0.2833 +8.1 i\,\num{e -10}}&{0.2835 +6.2 i\,\num{e -7}}\\ 
7&{0.3023 +0.0260 i}&{0.3023 +0.0260 i}&{0.3017 +0.0244 i}&{0.3022 +0.0260 i}\\ 
8&{0.3034 +1.7 i\,\num{e -13}}&{0.3034 +3.5 i\,\num{e -13}}&{0.2949 +1.5 i\,\num{e -14}}&{0.3034 +7.9 i\,\num{e -7}}\\ 
9&{0.3057 +3.8 i\,\num{e -6}}&{0.3057 +3.8 i\,\num{e -6}}&{0.3059 +3.5 i\,\num{e -6}}&{0.3057 +3.9 i\,\num{e -6}}\\ 
10&{0.3283 +3.5 i\,\num{e -8}}&{0.3283 +3.5 i\,\num{e -8}}&{0.3267 +3.2 i\,\num{e -8}}&{0.3283 +7.7 i\,\num{e -8}}\\ 
11&{0.3412 +4.1 i\,\num{e -13}}&{0.3412 +1.2 i\,\num{e -14}}&{0.3307 +5.0 i\,\num{e -15}}&{0.3412 +3.3 i\,\num{e -7}}\\ 
12&{0.3501 +0.1016 i}&{0.3501 +0.1016 i}&{0.3498 +0.1016 i}&{0.3501 +0.1017 i}\\ 
13&{0.3530 +1.9 i\,\num{e -5}}&{0.3530 +1.9 i\,\num{e -5}}&{0.3526 +1.5 i\,\num{e -5}}&{0.3529 +1.8 i\,\num{e -5}}\\ 
14&{0.3721 +3.7 i\,\num{e -12}}&{0.3721 +3.6 i\,\num{e -12}}&{0.3692 +3.1 i\,\num{e -12}}&{0.3721 +4.1 i\,\num{e -7}}\\ 
15&{0.3788 +3.7 i\,\num{e -13}}&{0.3788 +7.3 i\,\num{e -14}}&{0.3661 +1.8 i\,\num{e -14}}&{0.3788 +1.7 i\,\num{e -7}}\\ 
16&{0.3930 +0.0186 i}&{0.3930 +0.0186 i}&{0.3958 +0.0202 i}&{0.3930 +0.0186 i}\\ 
17&{0.4013 +2.6 i\,\num{e -8}}&{0.4013 +2.6 i\,\num{e -8}}&{0.4111 +2.5 i\,\num{e -10}}&{0.4013 +0.9 i\,\num{e -6}}\\ 
18&{0.4150 +0.9 i\,\num{e -10}}&{0.4150 +6.2 i\,\num{e -9}}&{0.4010 +4.4 i\,\num{e -12}}&{0.4150 +0.9 i\,\num{e -6}}\\ 
19&{0.4163 +1.7 i\,\num{e -9}}&{0.4163 +1.4 i\,\num{e -8}}&{0.4022 +2.8 i\,\num{e -8}}&{0.4163 +6.3 i\,\num{e -7}}\\ 
20&{0.4190 +4.0 i\,\num{e -5}}&{0.4190 +4.0 i\,\num{e -5}}&{0.4225 +4.4 i\,\num{e -5}}&{0.4190 +3.8 i\,\num{e -5}}\\ 
\bottomrule
\end{array}$

\caption{The first $20$ computed eigenfrequencies are displayed; the first column being the reference cavity, the second column is the ideal transformation and these agree very closely. The final two columns show small discrepancies; the third because of the layer discretization, and the fourth due to fictitious damping that mainly affects the imaginary parts. The frequency highlighted in red is related to a PML resonance.
}
\label{tab:freq comparison}
\end{table}

\begin{figure}
    \sisetup{output-exponent-marker=\ensuremath{\mathrm{e}}}
    \centering
    \begin{tabular}{p{0.01\textwidth}
        >{\centering}p{0.32\textwidth-2\tabcolsep - 1.666\arrayrulewidth}
        >{\centering}p{0.32\textwidth-2\tabcolsep - 1.666\arrayrulewidth}
        >{\centering}p{0.32\textwidth-2\tabcolsep - 1.666\arrayrulewidth}}
            & \textbf{Reference} & \textbf{Layer} & \textbf{NIM}
    \end{tabular}
    \\
        \raisebox{65pt}{\parbox{.02\textwidth}{\rotatebox{90}{\textbf{Mode 12}}}}
    \,\subfloat[$\complexqty{0.3501 + 0.1016i}{\hertz}$]{\includegraphics[width=0.32\textwidth,trim=0 5 5 40]{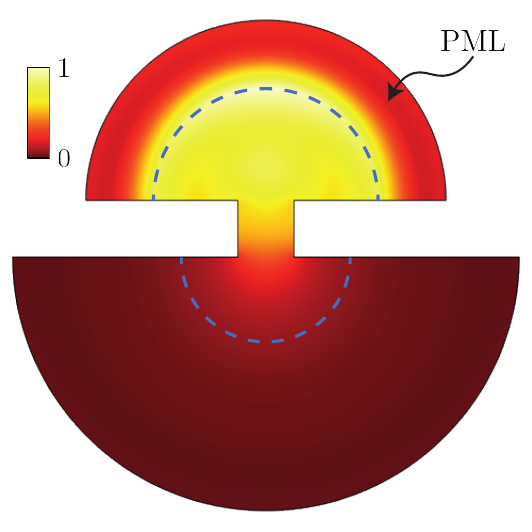}\label{fig:eigenmodes 2D-NIM-a}}
    \,\subfloat[$\complexqty{0.3501 + 0.1016i}{\hertz}$]{\includegraphics[width=0.32\textwidth,trim=0 -15 5 40]{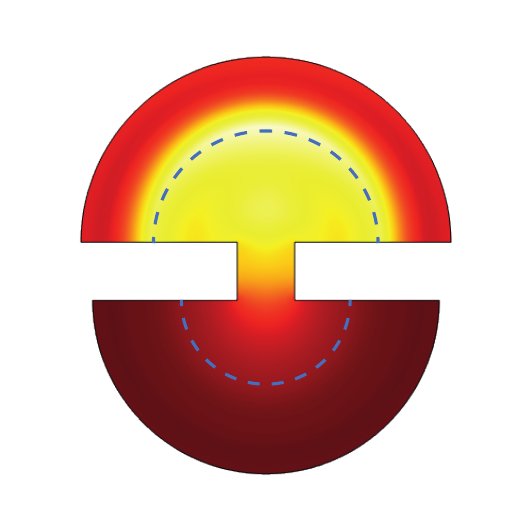}\label{fig:eigenmodes 2D-NIM-b}}
    \,\subfloat[$\complexqty{0.3498 + 0.1016 i}{\hertz}$]{\includegraphics[width=0.32\textwidth,trim=0 5 5 40]{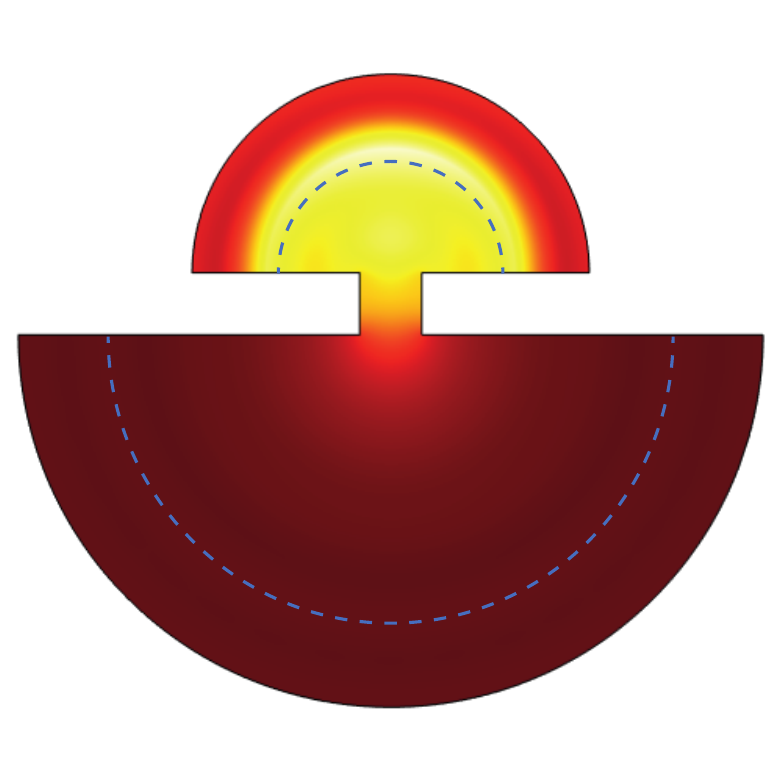}\label{fig:eigenmodes 2D-NIM-c}}
    \\
    \raisebox{65pt}{\parbox{.02\textwidth}{\rotatebox{90}{\textbf{Mode 16}}}}
    \,\subfloat[$\complexqty{0.3930 + 0.0186i}{\hertz}$]{\includegraphics[width=0.32\textwidth,trim=0 5 5 40]{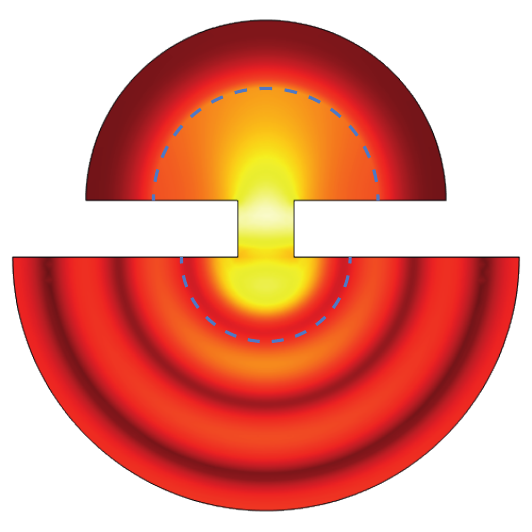}\label{fig:eigenmodes 2D-NIM-d}}
    \,\subfloat[$\complexqty{0.3930 + 0.0186i}{\hertz}$]{\includegraphics[width=0.32\textwidth,trim=0 -15 5 40]{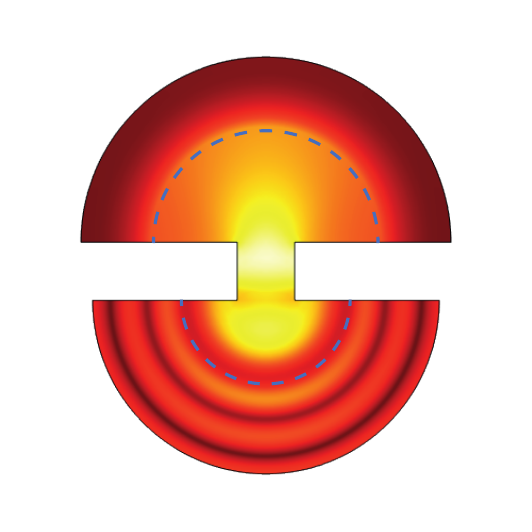}\label{fig:eigenmodes 2D-NIM-e}}
    \,\subfloat[$\complexqty{0.3958 + 0.0202 i}{\hertz}$]{\includegraphics[width=0.32\textwidth,trim=0 5 5 40]{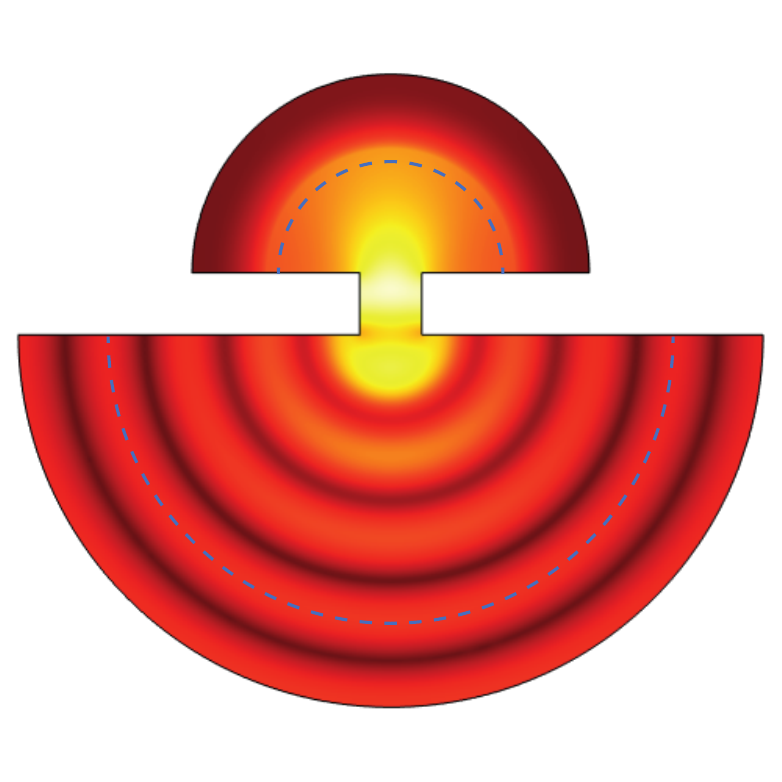}\label{fig:eigenmodes 2D-NIM-f}}
    \\
    \raisebox{65pt}{\parbox{.02\textwidth}{\rotatebox{90}{\textbf{Mode 17}}}}
    \,\subfloat[$(0.4013 +2.6 i\,\num{e -8})\unit{\hertz}$]{\includegraphics[width=0.32\textwidth,trim=0 5 5 40]{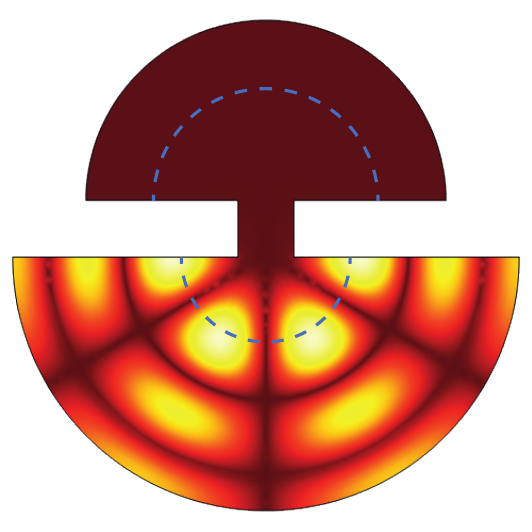}\label{fig:eigenmodes 2D-NIM-g}}
    \,\subfloat[$(0.4013 +2.6 i\,\num{e -8}) \unit{\hertz}$]{\includegraphics[width=0.32\textwidth,trim=0 -15 5 40]{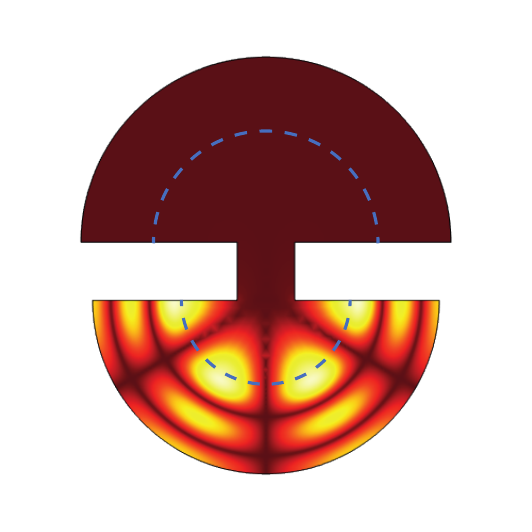}\label{fig:eigenmodes 2D-NIM-h}}   
    \,\subfloat[$(0.4111 +2.5 i\,\num{e -10})\unit{\hertz}$]{\includegraphics[width=0.32\textwidth,trim=0 5 5 40]{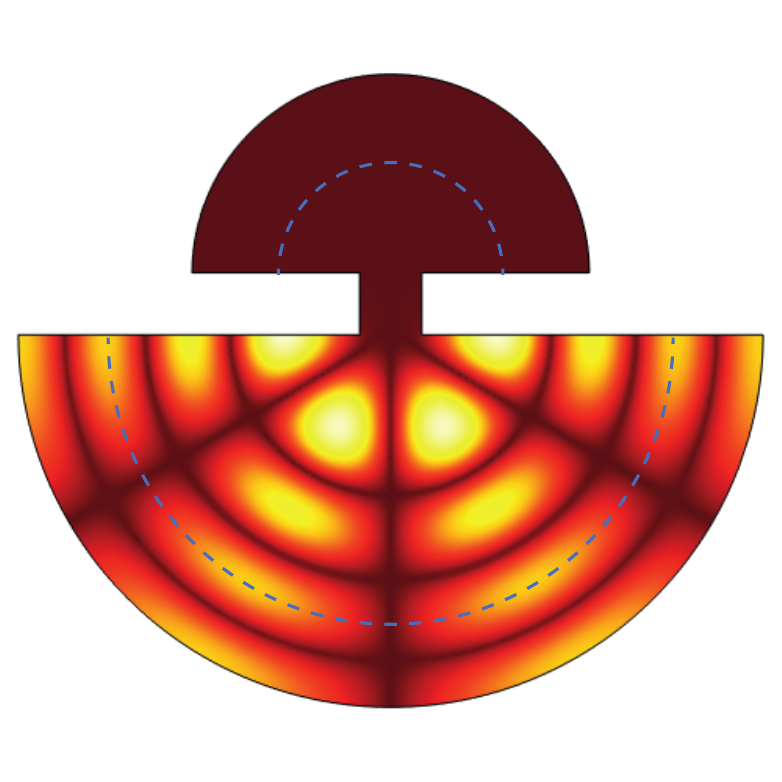}\label{fig:eigenmodes 2D-NIM-i}}
    \caption{The three columns show the absolute value of the pressure for three eigenmodes of the original cavity (left), the cavity cloaked with a layered metamaterial (centre), and the enlarged cavity equipped with a layer of NIM (right). \protect\subref{fig:eigenmodes 2D-NIM-a}, \protect\subref{fig:eigenmodes 2D-NIM-b} and \protect\subref{fig:eigenmodes 2D-NIM-c} show the twinning of a mode mainly characterized by leakage; \protect\subref{fig:eigenmodes 2D-NIM-d}, \protect\subref{fig:eigenmodes 2D-NIM-e} and \protect\subref{fig:eigenmodes 2D-NIM-f} show a trapped resonance with some leakage; \protect\subref{fig:eigenmodes 2D-NIM-g}, \protect\subref{fig:eigenmodes 2D-NIM-h} and \protect\subref{fig:eigenmodes 2D-NIM-i} show a pure trapped mode inside the cavity. The dashed blue lines highlights the PML, the undeformed domain, and the metamaterial layers.
    }
    \label{fig:eigenmodes 2D-NIM}
\end{figure}

As an example, some eigenmodes are displayed in Figure~\ref{fig:eigenmodes 2D-NIM} and notably we do not show the ideal case as the eigenvalues are identical to the reference case, in so far as digits shown, and the eigenfields are visually indistinguishable from each other. Since the precise comparison of the fields is not straightforward, a measure of the twinning is given by applying the so-called \textit{Modal Assurance Criterion} (MAC) \cite{allemang2003modal} which allows us to compare two eigenmodes by computing the following matrix:
\begin{equation}
    MAC^{ideal}_{ij}\coloneqq\frac{\left|\int_{\Sigma_f} p^{}_{i,ideal}\,p^*_{j,ref}\right|^2}{\int_{\Sigma_f} p^{}_{i,ideal}\,p^*_{i,ideal}\,\int_{\Sigma_f} p^{}_{j,ref} p^*_{j,ref}},
\end{equation}
where $\Sigma_f\equiv\sigma_f$ is the portion of the geometry that does not undergo any transformation, $p^*$ is the complex conjugate of $p$, and the subscripts $ref$, $ideal$, $layer$, and $NIM$ indicate respectively the reference cavity, the ideally twinned cavity, and the two cavities equipped with the layered and the NIM gratings. The indexes $i$ and $j$ specify the eigenmodes under comparison. Note that the MAC matrix is real valued and it assumes unitary values if and only if the two fields under comparison are identical inside $\sigma_f$, except for an arbitrary scaling factor. So one can state that the twinning is satisfactory for all eigenmodes if the main diagonal terms of the $MAC$ matrix are close to the unity. 

\begin{figure}
    \centering
    \subfloat[]{
    \includegraphics[height=4.5cm,trim=29 8 108 0,clip]{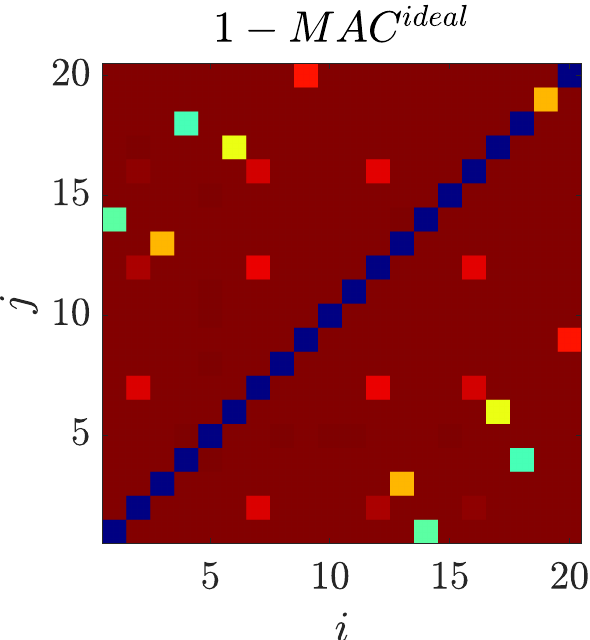}
    \label{fig:MAC chart-a}}
    \subfloat[]{
    \includegraphics[height=4.5cm,trim=29 8 108 0,clip]{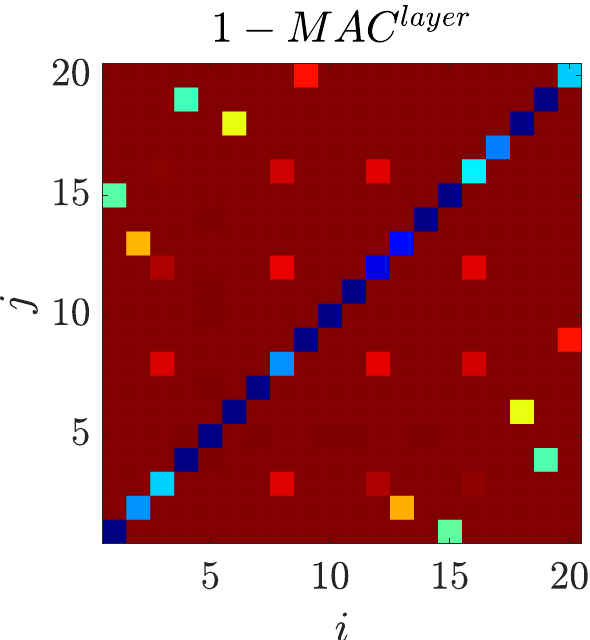}
    \label{fig:MAC chart-b}}
    \subfloat[]{
    \includegraphics[height=4.5cm,trim=29 8 44 0,clip]{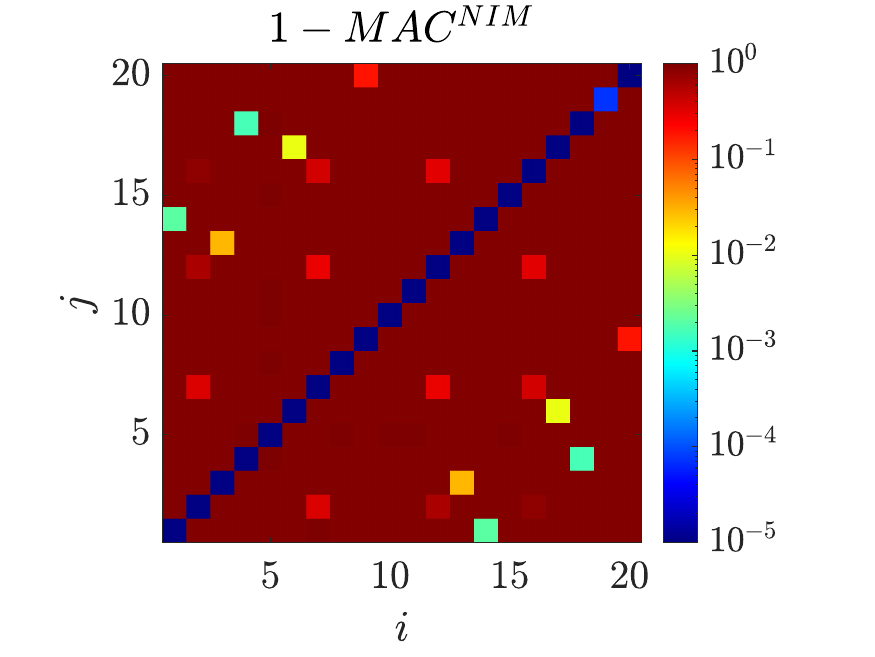}
    \label{fig:MAC chart-c}}
    \caption{The computed eigenmodes of the original cavity are compared with \protect\subref{fig:MAC chart-a} the ideal,\protect\subref{fig:MAC chart-b} the layer, and the \protect\subref{fig:MAC chart-c} NIM twin cavities. We use a logarithmic scale to allow the twinning to be more easily seen; more accurate twinning corresponds to smaller values.  
    }
    \label{fig:MAC chart}
\end{figure}

Thus the \textit{ideal}, the \textit{layer}, and the \textit{NIM} twin cavities are compared to the reference one.
Figure~\ref{fig:MAC chart} displays the matrices $1-MAC^{ideal}_{ij}$, $1-MAC^{layer}_{ij}$, and $1-MAC^{NIM}_{ij}$ using a logarithmic colour scale, such that good twinning is highlighted by values close to $0$. Figure~\ref{fig:MAC chart-a}, relative to the \textit{ideal} twin, shows a complete agreement for each and every mode, except the 19\textsuperscript{th}. This mode is very localized in the transformed region $\sigma_c$, so the agreement cannot be captured by comparing $\sigma_f$ only.
Figure~\ref{fig:MAC chart-b} proves that the layered arrangement approximates the ideal behaviour because most of the modes show excellent agreement with the reference ones. Lastly, Figure~\ref{fig:MAC chart-c} shows a weaker match, due to the fictitious damping.
Some of the off-diagonal terms are close to $0$ even if the corresponding modes have very different frequencies. Indeed, those modes have a similar shape inside the test region $\sigma_f$, but since they are far in frequency, they can be easily disregarded.

\begin{figure}
    \centering
    \subfloat[]{\includegraphics[height=5.3cm,trim=20 0 20 0,clip]{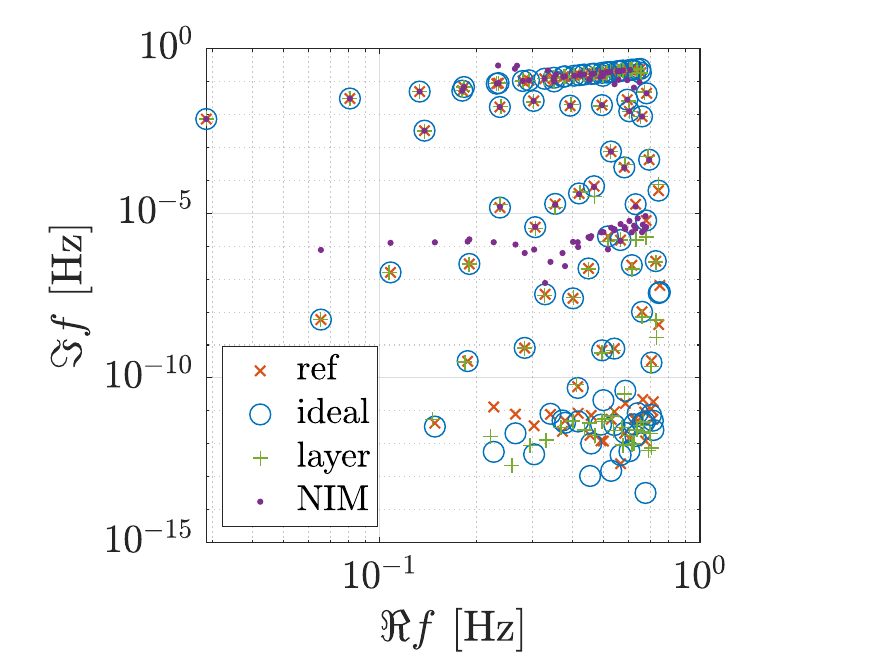}\label{fig:MAC freq-a}}
    \,\,
    \subfloat[]{\includegraphics[height=5.3cm,trim=0 0 18 0,clip]{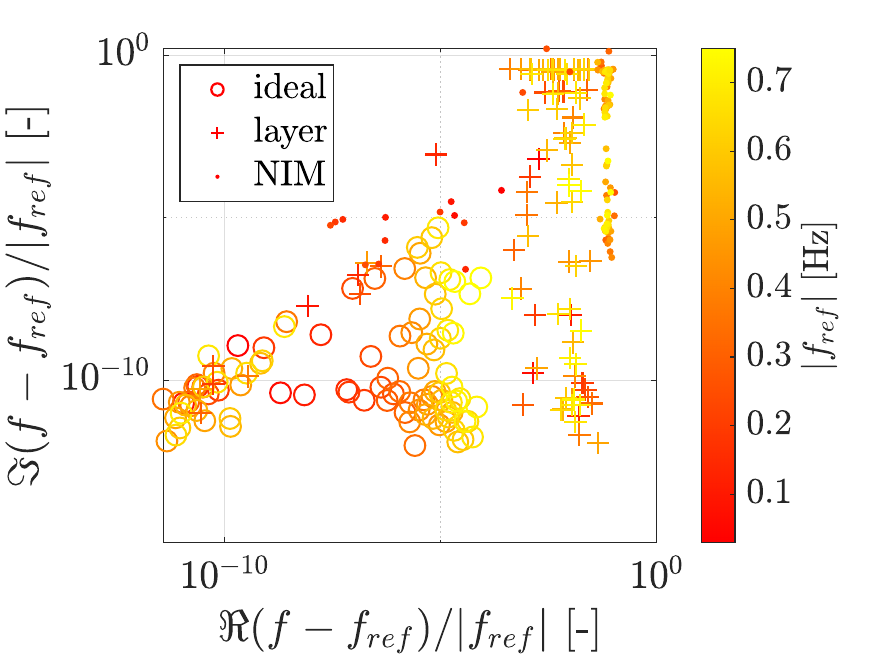}\label{fig:MAC freq-b}}
    \caption{The first $100$ eigenfrequencies of the four cavities are compared in \protect\subref{fig:MAC freq-a} and the relative errors with respect to the reference are shown in \protect\subref{fig:MAC freq-b}. On the complex plane, we can distinguish an oblique straight line collecting the resonances induced by the PML and a cloud of points close to the real axis that contains the trapped cavity resonances.
    }
    \label{fig:MAC freq}
\end{figure}

Finally, the spectra of the four cavities are compared by computing the first $100$ eigenfrequencies. Figure~\ref{fig:MAC freq-a} shows that they superpose in the complex plane and the relative errors in Figure~\ref{fig:MAC freq-b} are small: below \SI{1e-5}{} for the ideal twin, below \SI{1e-2}{} for the layered twin, and below \SI{5e-2}{} for the NIM twin. Thus we obtain, numerically, that the spectrum is matched for both real and imaginary parts and hence that our design methodology is validated for open systems.

\section{Flattening a blazed grating}
\label{sec:metasurf}
\begin{figure}
    \centering
    \includegraphics[width=0.85\textwidth]{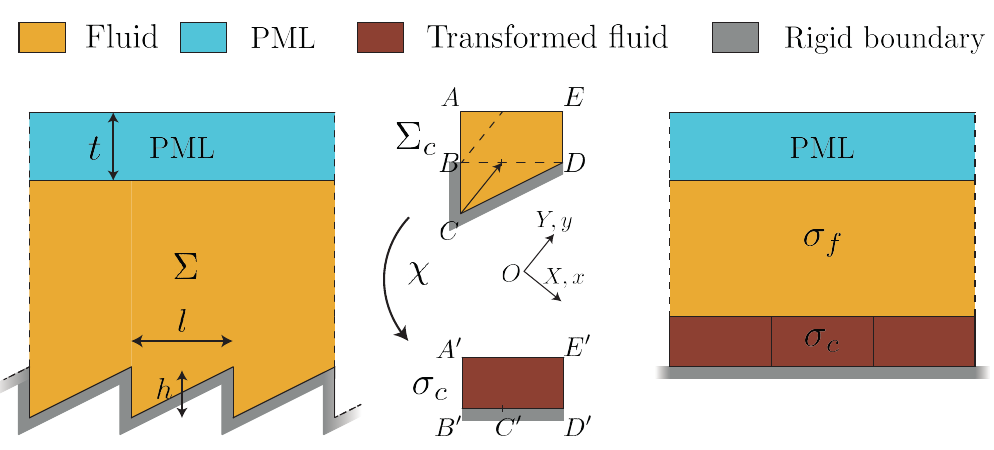}
    \caption{Sketch of the metasurface and transformation. Left: the saw-tooth grating described by its depth $h$ and period $l$; middle: a portion of the domain $\Sigma_c$ is transformed to a rectangle called $\sigma_c$ using a projection with respect to the rotated reference systems $XOY=xOy$; right: flattened metasurface having the same acoustic behaviour as the saw-tooth grating. $l=\SI{1}{\meter}$, $h=l/2$, $t=2l$.
    }
    \label{fig:metasurf geom}
\end{figure}

We now illustrate our approach for twinning open cavities by considering a class of diffraction gratings. Structured surfaces, such as the saw-tooth one depicted in Figure~\ref{fig:metasurf geom}, usually called \textit{blazed} or \textit{echelette} gratings, are widely used in electromagnetism, for instance in high-resolution spectroscopy and spectrometers \cite{Probst2017,Barnard1993}, for their ability to diffract incident light into a given direction \cite{loewen2018diffraction}.
We now create a metasurface having the same behaviour as the echelette grating, but characterised by a completely flat profile, and use the formalism of acoustics.

Each unit cell of the echelette is considered as an open cavity, and a twinned cell is defined to enable us to build a flat metasurface thinner than the reference grating. Finally, we test the twinning showing a complete agreement between the dispersion diagrams of the grating and the flat metasurface.
Using the periodicity along the horizontal direction the acoustic behaviour of the metasurface, displayed in Figure~\ref{fig:metasurf geom}, is analysed by considering a single unit cell of width $l$ and imposing the Floquet-Bloch periodicity on the two vertical cell boundaries. 
We now focus on the transformation applied to the geometry of each cell such that the saw-tooth metasurface is mapped into a flat one.

As in section \ref{sec:metasurf}, the elementary cell $\Sigma$ (filled with a fluid having $\rho_0=\SI{1}{\kilo\gram\per\meter^3}$ and $\kappa=\SI{1}{\pascal}$) is partitioned into two domains: $\Sigma_c$ subjected to the transformation and $\Sigma_f$ kept unaltered.
The transformation we use involves the entire triangular cavity and an arbitrary portion of the fluid outside which is necessary in order to avoid singularities. The rigid boundary $\Gamma$ is continuously mapped onto a flat boundary; thus, for example, the point $C$ cannot be mapped on $B$ or $D$, but is mapped to a point between them.
The transformation we choose projects the point $C$ on the line $BD$ along the bisector of the angle $\hat{BCD}$, as shown in Figure~\ref{fig:metasurf geom}.

This transformation is not unique and one can transform by stretching the domain in many different ways, controlling the amount of stretch imposed to the geometry or even imposing a space folding.
This flexibility is potentially valuable in practical terms as according to the precice choice, the transformed properties can scale towards values smaller or greater than those of the fluid, and can be influenced by the availability of materials for implementation.

\begin{figure}
    \centering
    \subfloat[]{\includegraphics[width=0.485\textwidth]{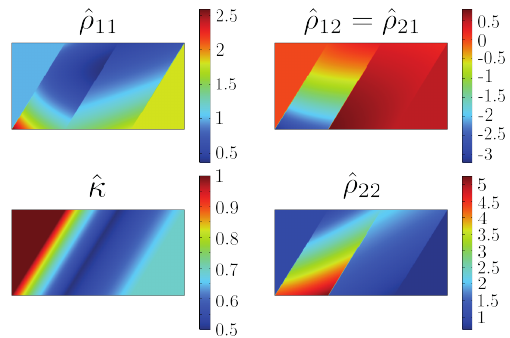}\label{fig:metasurface prop-a}}
    \quad
    \subfloat[]{\includegraphics[width=0.485\textwidth]{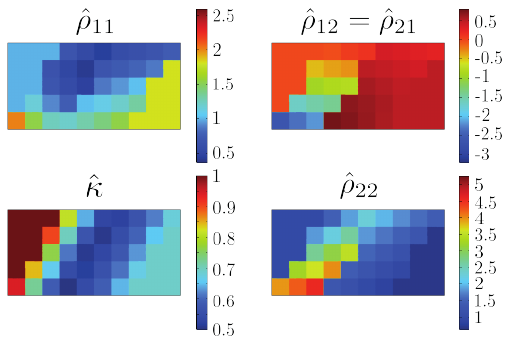}\label{fig:metasurface prop-b}}
    \caption{Transformed properties normalised with respect to the fluid properties $\kappa_0$ and $\rho_0$. The transformation changes the material properties according to the stretch imposed. Values  are displayed in colour scale, for the ideal \protect\subref{fig:metasurface prop-a} and discretized \protect\subref{fig:metasurface prop-b} twins.
    Note that the transformation keeps the properties unaltered in a triangle close to the upper left vertex since the curves $Y_0$ and $Y_1$ are superimposed along $AB$.
    } 
    \label{fig:metasurface prop}
\end{figure}

For the transformation we take the orthonormal rotated reference systems $XOY\equiv xOy$ displayed in Figure~\ref{fig:metasurf geom}, where one axis is parallel to the bisector of the angle $\hat{BCD}$; this simplifies the analytical expression of the transformation. Following \cite{lenz2023transformation} we define three curves $Y_0(X)$, $Y_1(X)$ and $Y_2(X)$ that correspond to the lines $ACD$, $ABD$, and $AED$ respectively.
The transformation $\bs\chi$ is such that the points between $Y_0$ and $Y_2$ are linearly mapped to points between $Y_1$ and $Y_2$. That is:
\begin{equation}
\begin{aligned}
    x =& X,
    \\
    y =& \alpha(X)Y + \beta(X),
\end{aligned}
\end{equation}
where $\alpha(X)\coloneqq \frac{Y_2(X)-Y_1(X)}{Y_2(X)-Y_0(X)}$, $\beta(X)\coloneqq Y_1(X)-Y_0(X)\alpha(X)$.
The deformation gradient is then given by:
\begin{equation}
    \bb J =
    \begin{bmatrix}
        1 & 0
        \\
        \alpha' Y+\beta' & \alpha
    \end{bmatrix},
\end{equation}
and the transformed properties $\bs\rho_c$ and $\kappa_c$ are computed according to Eq.~\eqref{eq:transf prop}. The tensor $\bs\rho_c$ is related to the rotated reference system $xOy$, thus it has been rotated such that its components are referred to the original Cartesian system; the plots of Figure~\ref{fig:metasurface prop-a} show the properties obtained on a colour scale.

\begin{figure}
    \centering
    \subfloat[]{\includegraphics[width=0.45\textwidth,trim=40 0 50 0,clip]{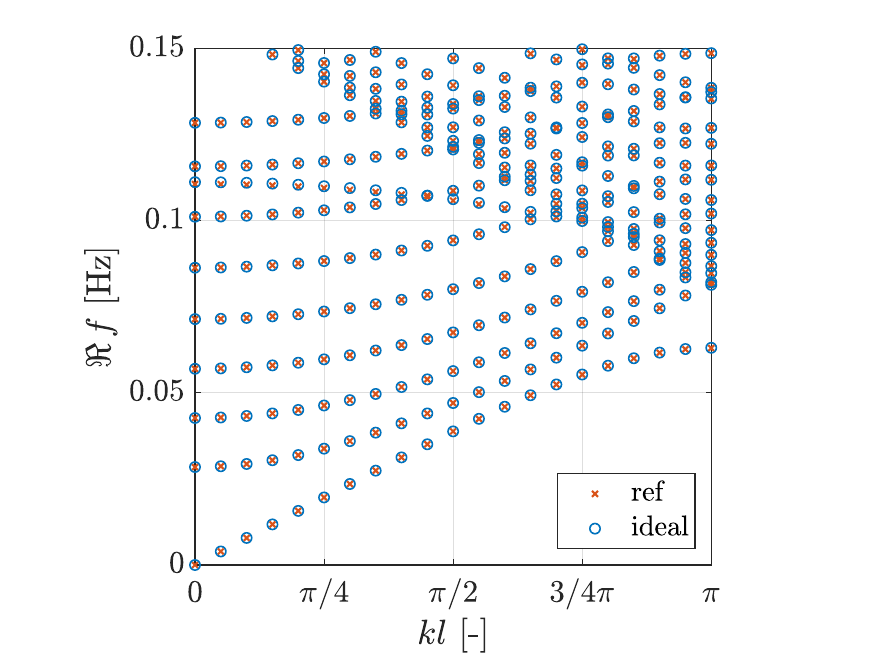}\label{fig:dispersion-a}\qquad}
    \subfloat[]{\includegraphics[width=0.45\textwidth,trim=40 0 50 0,clip]{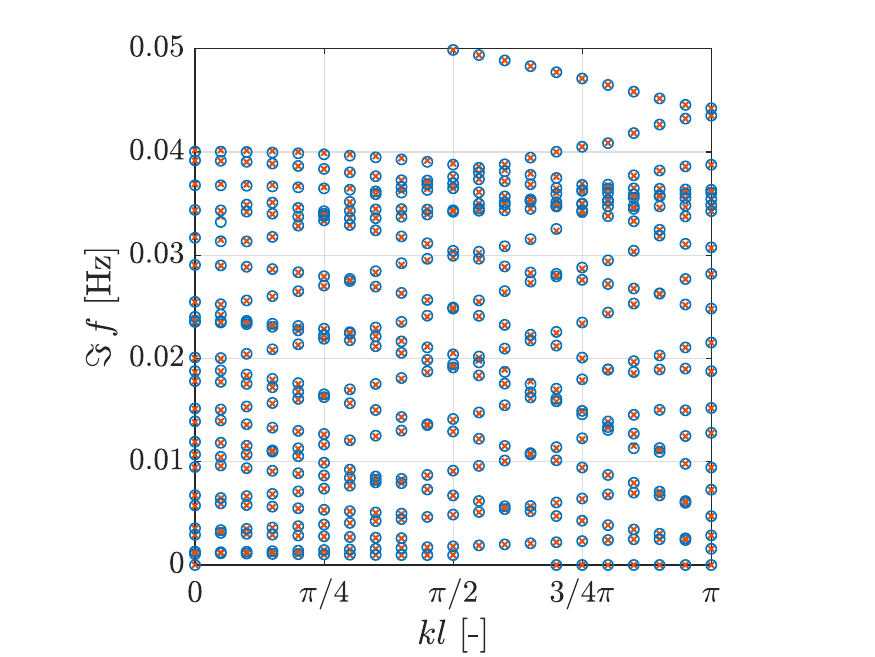}\label{fig:dispersion-b}\qquad}
    \caption{The two dispersion diagrams of the metasurface and its ideal twin superimposed; \protect\subref{fig:dispersion-a} real and \protect\subref{fig:dispersion-b} imaginary parts of the frequency.
    }
    \label{fig:dispersion}
\end{figure}

Numerical simulations are performed to find the dispersion curves, i.e. the eigenspectra, that relate frequency to Bloch wavenumber, $k$, and these use the original and transformed unit cells in the finite element simulations that are augmented with a cartesian PML that truncates the semi-infinite medium. From periodicity the dispersion diagrams only require variation of the wavenumber $k$ in the first Brillouin zone $[0,\pi/l]$.
Figure~\ref{fig:dispersion} compares the dispersion diagrams of the twinned cavities: both real and imaginary parts of the frequency show a very good agreement, indicating the ability to twin all the dispersion properties for these periodic structures. We note that a similar approach was implemented in \cite{chatzopoulos2022cloaking} to quantify the cloaking efficiency of carpet cloaks. However, dispersion diagrams  were an aside in a study addressing scattering problems.

Turning now to creating an effective medium approximation using discrete elements made from materials with homogeneous properties. Noting that significant inhomogeneity can arise in the transformed medium, so the layered arrangement adopted in Section~\ref{sec:num sol} is not possible due to the rapid spatial variations, we introduce an intermediate discretization: a graded metamaterial, constructed from simple components, is designed by dividing the transformed domain into small cells with homogeneous properties, such that each cell can be constructed using, e.g., a layered medium with different orientations with anisotropic effective properties. \\

\begin{figure}
    \centering
    \subfloat[]{\includegraphics[height=5.5cm,trim= 10 0 30 0,clip]{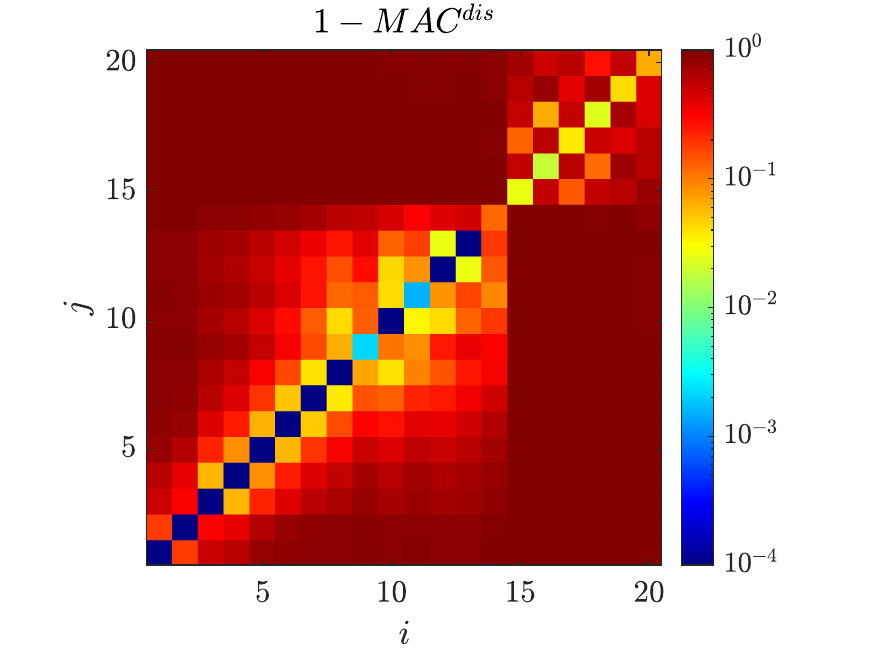}\label{fig:discrete metasurf-a}} \,
    \subfloat[]{\includegraphics[height=5.5cm,trim= 10 0 30 0,clip]{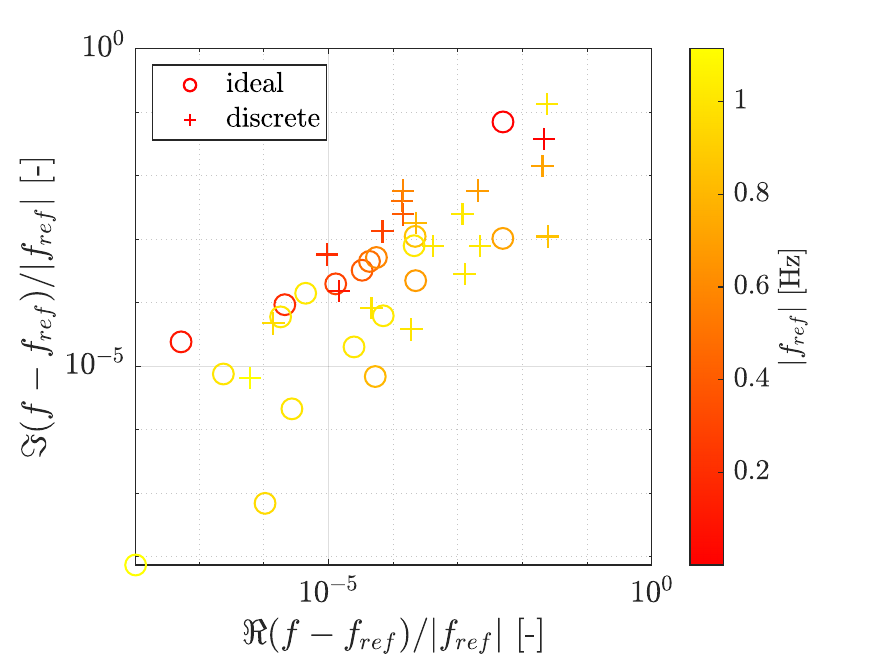}\label{fig:discrete metasurf-b}}
    \caption{\protect\subref{fig:discrete metasurf-a} MAC comparing the discretized twin against the reference echelette; \protect\subref{fig:discrete metasurf-b} errors of the complex valued eigenfrequencies for the ideal and the discrete twins.
    }
    \label{fig:discrete metasurf}
\end{figure}

To evaluate this discretized approach, the transformed domain is divided into square cells of size $l/10$ and the properties are considered piecewise constant such that each pixel has the values given by $\bs\rho_c(\bb x)$ and $\kappa_c(\bb x)$ evaluated in its centre. The discretized arrangement is showed in Figure~\ref{fig:metasurface prop-b}. Numerical simulations enable us to evaluate the twinning numerically and we show results for the Floquet-Bloch ansatz with $k=\SI{0}{\radian\per\meter}$ and the eigenmodes and the eigenfrequencies of the twins are compared through the MAC and the errors shown in Figure~\ref{fig:discrete metasurf}. In particular, Figure~\ref{fig:discrete metasurf-a} highlights that eigenmodes corresponding to higher frequencies are more affected by the discretization than the low frequency modes; higher frequencies are associated with shorter wavelengths that are more sensitive to a discrete arrangement with similar lengthscales. 
Figure~\ref{fig:discrete metasurf-b} shows the errors introduced on the eigenmodes by the ideal and the discrete twinning; while the former comes from numerical approximations only, the latter mainly suffers from the coarse discretization used.

\section{Conclusions}
\label{sec:conclusions}

The scheme validated through this work suggests a viable way to design the shape of twin open cavities, i.e.\ those cavities intended to resonate with the same spectrum of a reference cavity, but having a different geometry. We have chosen examples to display the versatility of the approach, i.e.\ a resonant Helmholtz cavity, which is a good test of the approach having both trapped and leaky modes to reproduce and then a case from periodic systems, which is non-resonant but now has sharp corners. The upshot is that one could take designs from the literature that use constant parameter acoustic fluids as the reference and then transform then to other more convenient geometries. The resultant cavity/surface would have some region with anisotropic fluid within it but, as demonstrated, this could be replaced by layered or discrete homogeneous media as effective replacements.

We show that unbounded domains can be twinned by applying transformation theory: this reveals multiple ways of resizing an open cavity, for instance, the inner wall of a small cavity is covered with an \textit{ad hoc} layered medium so that it has a spectrum identical to a larger one (or a smaller one for a space folding transform). Both the discrete and the continuous branches of the complex valued spectrum are restored within very good tolerance regardless of the quality factor of the eigenmodes. \\ 
We also chose a more challenging case consisting in flattening a blazed grating, and the method is shown to be robust even with strong inhomogeneities. The accuracy of the twinning remains within a good tolerance even when a coarse discretization oriented to realise the metasurface is taken into account. Via the Floquet-Bloch theorem, the assessment is performed comparing the dispersion diagrams of the grating and its twins. 

In the context of metasurfaces based around resonators we anticipate that the mathematical approach taken here will allow thinner surfaces to be designed that have low frequency resonators taking up less space. The twinning works almost perfectly for an ideal anisotropic fluid and still provides isospectrality even when an effective medium, made of homogeneous layers, is used to mimic the anisotropic fluid. Our theoretical and numerical results hold in the context of anti-plane shear elasticity and transverse electromagnetism. This provides confidence that such twinned cavities could be fabricated and motivates further experimental work.




\bibliographystyle{RS}
\bibliography{main}

\end{document}